\documentclass[sigconf]{acmart}
\usepackage{graphicx}
\usepackage{amsmath}
\usepackage{booktabs}
\usepackage{algorithm}
\usepackage{algorithmic}
\usepackage{amsfonts}
\usepackage{multirow}
\usepackage{makecell}
\usepackage{subfigure}
\usepackage{color}
\usepackage{bm}
\usepackage{epstopdf}
\usepackage{url}
\usepackage[cal=cm]{mathalfa}
\usepackage{balance}
\usepackage{threeparttable}
\usepackage{lipsum}
\usepackage{enumitem}

\AtBeginDocument{%
  \providecommand\BibTeX{{%
    \normalfont B\kern-0.5em{\scshape i\kern-0.25em b}\kern-0.8em\TeX}}}

\author{Xinchen Luo, Jiangxia Cao, Tianyu Sun$^*$, Jinkai Yu$^*$, Rui Huang$^*$, Wei Yuan, Hezheng Lin, Yichen Zheng, Shiyao Wang, Qigen Hu, Changqing Qiu, Jiaqi Zhang, Xu Zhang, Zhiheng Yan, Jingming Zhang, Simin Zhang, Mingxing Wen, Zhaojie Liu, Kun Gai, Guorui Zhou}
\thanks{* Authors who did really hard work in the deployment.}
\affiliation{
  \institution{Kuaishou Technology, Beijing, China}
 \country{\{luoxinchen, caojiangxia, suntianyu06, yujinkai, huangrui06, yuanwei05, linhezheng, zhengyichen, wangshiyao08, huqigen03, qiuchangqing, zhangjiaqi15, zhangxu20, yanzhiheng, zhangjingming, zhangsimin03, wenmingxing, zhaotianxing, zhouguorui\}@kuaishou.com; kun.gai@qq.com}
}

\begin{document}
\title{QARM: Quantitative Alignment Multi-Modal Recommendation at Kuaishou}
\renewcommand{\shorttitle}{QARM}

\begin{abstract}
In recent years, with the significant evolution of multi-modal large models, many recommender researchers realized the potential of multi-modal information for user interest modeling.
In industry, a wide-used modeling architecture is a cascading paradigm: (1) first pre-training a multi-modal model to provide omnipotent representations for downstream services; (2) The downstream recommendation model takes the multi-modal representation as additional input to fit real user-item behaviours.
Although such paradigm achieves remarkable improvements, however, there still exist two problems that limit model performance:
(1) \textbf{Representation Unmatching}: The pre-trained multi-modal model is always supervised by the classic NLP/CV tasks, while the recommendation models are supervised by real user-item interaction. As a result, the two fundamentally different tasks' goals were relatively separate, and there was a lack of consistent objective on their representations;
(2) \textbf{Representation Unlearning}: The generated multi-modal representations are always stored in cache store and serve as extra fixed input of recommendation model, thus could not be updated by recommendation model gradient, further unfriendly for downstream training.

Inspired by the two difficulties challenges in downstream tasks usage, we introduce a quantitative multi-modal framework to customize the specialized and trainable multi-modal information for different downstream models.
Specifically, we introduce two insightful modifications to enhance above framework:
(1) \textbf{Item Alignment} to transform the original multi-modal representations to match the real user-item behaviours distribution.
(2) \textbf{Quantitative Code} to transform the aligned multi-modal representations to trainable code ID for downstream tasks.
We conduct detailed experiments and ablation analyses to demonstrate our QARM effectiveness.
Our method has been deployed on Kuaishou’s various services, serving 400 million users daily.

\end{abstract}



\maketitle

\section{Introduction}
Kuaishou, is one of the largest short-video and live-streaming platform in China.
As a new type of information-sharing media, Kuaishou attracts a lot of attention and accumulates a large number of users to watch/create short-videos, and even shopping goods after watching some online-shopping and advertising short-videos or live-streamings.
To find the most interesting short-videos content from billions of short-videos pool and provide satisfied experience for our users, a strong recommender system (RecSys) is a cornerstone to support Kuaishou business~\cite{home,crossmoment}.
Generally, to obtain a powerful recommendation model, the common wisdom always holds the idea that the model should be learning from the massive real-time user-item interaction data with a large number of hand-craft model input features.
In past years, many recommendation engineers/researchers proposed several milestones work to elaborate the ID-based features to support model input features, such as cross ID features (e.g., FM~\cite
{fm}, DCN~\cite{dcn}), list-wise ID features (e.g. DIN~\cite{din}, TWIN~\cite{twin}).
In recent years, with the significant evolution of multi-modal large models (e.g., GPTs~\cite{gpt2}), many recommender engineers/researchers realized the potential of multi-modal information in recommender area, to understand the item certain semantic signal to make more smart recommendation.
Especially at the platform at Kuaishou, the short-video and live-stream are highly integrated multi-modal media, it is difficult to fully understand them by assigning an ID embedding only.

\begin{figure*}[t]
  \centering
  \includegraphics[width=18cm,height=6cm]{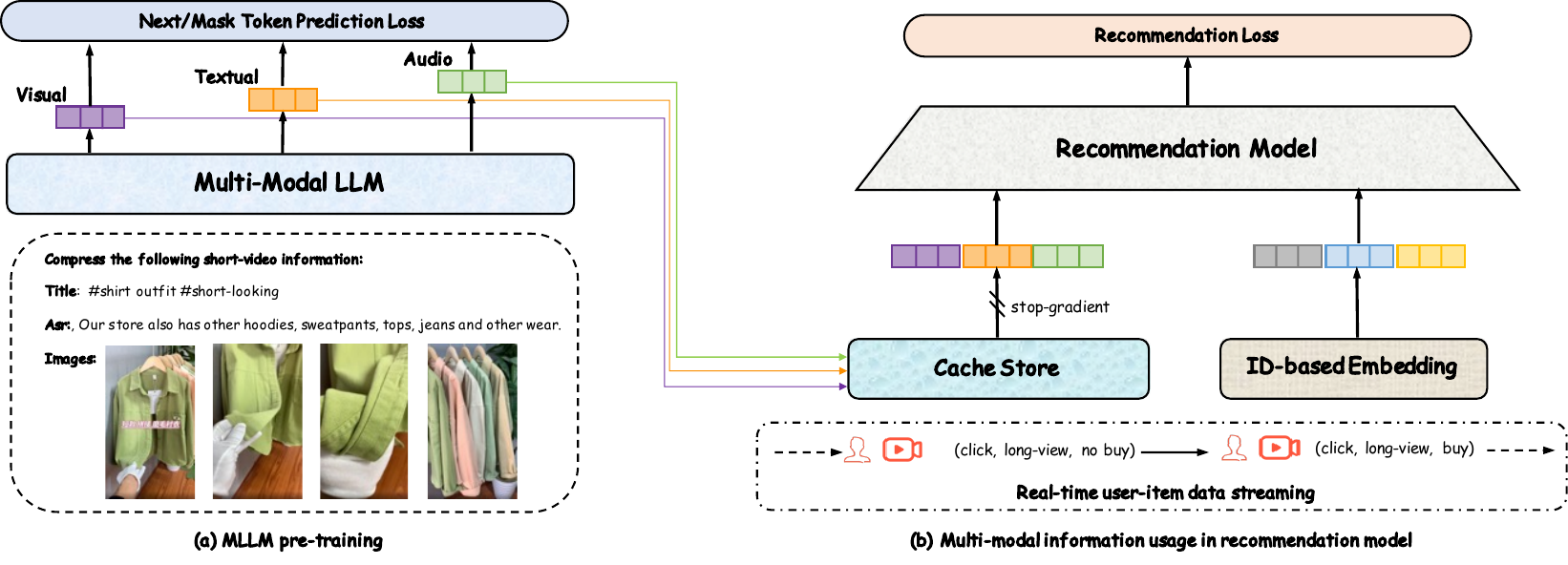}
  \caption{(a) Representation Unmatching: multi-modal features are obtained through down-streaming task-agnostic self-supervised tasks like image-text matching. (b) Representation Unlearning: generated multi-modal representation is always only served as additional fixed static input feature for recommendation model.}
  \label{fig:intro}
\end{figure*}

However, although powerful, the multi-modal large language models (MLLMs) are notorious for their tremendous computation cost in training and inference, considering the massive online requests for industrial recommendation services, it is impractical to directly add a large multi-modal module into recommendation model~\cite{simtier}.
To alleviate the computation pressure, to our knowledge, many companies apply a two-step deployment solution to incorporate such MLLM-based semantic information for recommendation model (As shown in Figure~\ref{fig:intro}):
(1) \textbf{Pretraining} a MLLM to compress the items' text, visual and audio information as an omnipotent representation.
Next utilize the pre-trained MLLM to produce item representations, and save them in a cache store to provide the world semantic knowledge for downstream models.
(2) According to the training sample information, the downstream models could \textbf{fetch} the corresponding necessary multi-modal representation as a part of input features, to enhance the model's prediction ability.
At Kuaishou, our recommendation models are also equipped with such deployment solution, and achieve remarkable online A/B gains in different businesses, such as online-shopping and advertise short-video and live-streaming recommendation.
Nevertheless, such non-end-to-end framework has two obvious problems limit model performance:

\textbf{Representation Unmatching}: 
Common multi-modal features are obtained through self-supervised tasks like image-text matching\cite{clip, blip2}, while ID-based features use user interaction history as supervision signals\cite{twin, din}. These differences make it challenging to unify multi-modal information and recommendation knowledge in downstream training. Therefore, a question is that: \textit{Can we enhance the multi-modal
representation consistency for downstream task?}

\textbf{Representation Unlearning}: 
In practice, the newly added multimodal features do not update with the training of the recommendation system\cite{simtier}.
Nevertheless, for the others discrete ID-based features (e.g., user ID, item ID), our recommendation model could assign a corresponding embedding spaces to them for end-to-end optimization with the real-time user-item interaction data.
Consequently, the static multi-modal representations are easily limit the model fitting ability and obstacles model convergence.
As a result, there rise another question: \textit{Can we allow multi-modal representations optimized in end-to-end manner?}
Motivated by the two difficulties challenges in downstream tasks usage, in this paper, we present our effective and efficient solutions for multi-modal information enhancement, the \underline{Q}uantitative \underline{A}lignment Multi-Modal \underline{R}eco\underline{m}mendation, termed as \textbf{QARM}.
Specifically, our QARM consists of two major processes to answer the above questions, the item alignment mechanism to enhance the representation consistency, and the quantitative code mechanism to generate learnable code ID for downstream tasks.

\textbf{Item Alignment mechanism}: 
To alleviate the first representation unmatching challenge and maximize the representation consistency ability, we consider fine-tuning the pre-trained multi-modal model in a customized manner.
The reason is that different businesses have different characteristics, and the downstream task expected multi-modal representation should express corresponding business characteristics.
For example, the causal relationship between different categories of goods is beneficial for online-shopping short-videos services while the same category item relationship is more important for usual short-videos recommendation.
Therefore, such fine-tuning paradigm should be customized for each type of downstream business.
To implement it, we insert a pre-order item alignment mechanism to fine-tune the multi-modal model with corresponding business data, to encourage MLLM representation could reflect the real business user-item interaction pattern.

\textbf{Quantitative Code mechanism}: To overcome the second frozen representation unlearning challenge for full multi-modal information adaptation, inspired by the code hashing~\cite{leskovec2020mining} and straight-through estimator~\cite{van2017neural,singh2024better} idea, we consider generating the Semantic IDs for the down-stream tasks.
Specifically, after obtaining the fine-tuned multi-modal representation, we propose a simple-but-effective heuristics residual K-means algorithm to obtain a quantization codebook.
Once the quantization codebook is trained, we next freeze the codebook and use it to measure the fine-tuned multi-modal representation to calculate corresponding Semantic IDs. 
Finally, in downstream recommendation model training, we assign corresponding embedding space for Semantic IDs for end-to-end training with our real user-item interaction data.

\begin{figure*}[t]
  \centering
  \includegraphics[width=18cm,height=4cm]{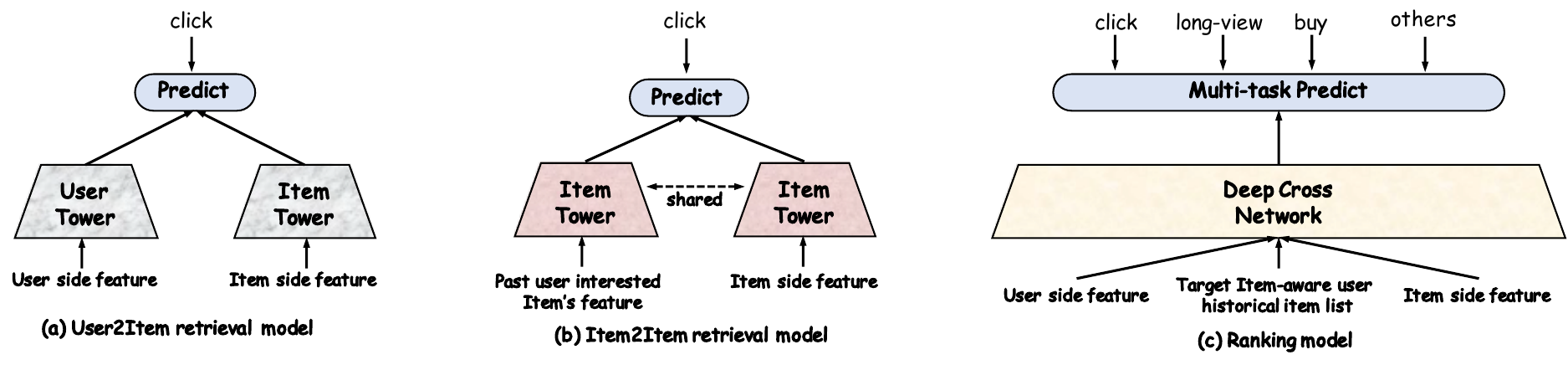}
  \caption{The industry RecSys architectures: (a)User2Item/(b)Item2Item retrieval model, and (c) User-Item ranking model.}
  \label{fig:background}
\end{figure*}

In summary, the main contributions of this paper are as follows:
\begin{itemize}[leftmargin=*,align=left]
\item We present QARM, a novel method to overcome two major limitations of industrial multi-modal information applications. Specifically, the item alignment mechanism generates consistent multi-modal representations for downstream business, while the quantitative code mechanism further compresses them as learnable semantic IDs for end-to-end training.
\item We conduct extensive offline experiments and ablation studies to show our QARM component effectiveness. We also conduct online experiments, which find that our QARM contributes Revenue +9.704\% in advertising, and GMV+2.296\% in online-shopping.
\item QARM has been widely validating various services at Kuaishou since 2024 March, supporting 400 million active users daily.
\end{itemize}

\section{Methodology}
In this section, we explain our QARM components and the total deployment workflow.
Before going on, we first retrospect the background of industrial RecSys, including the feature engineering, and the training paradigm of different models used in different stages.
Next, we dive into QARM, to express the item alignment mechanism for representation unmatching problem and the quantitative code mechanism for representation unlearning problem in detail.

\begin{figure*}[t]
  \centering
  \includegraphics[width=18cm,height=4.7cm]{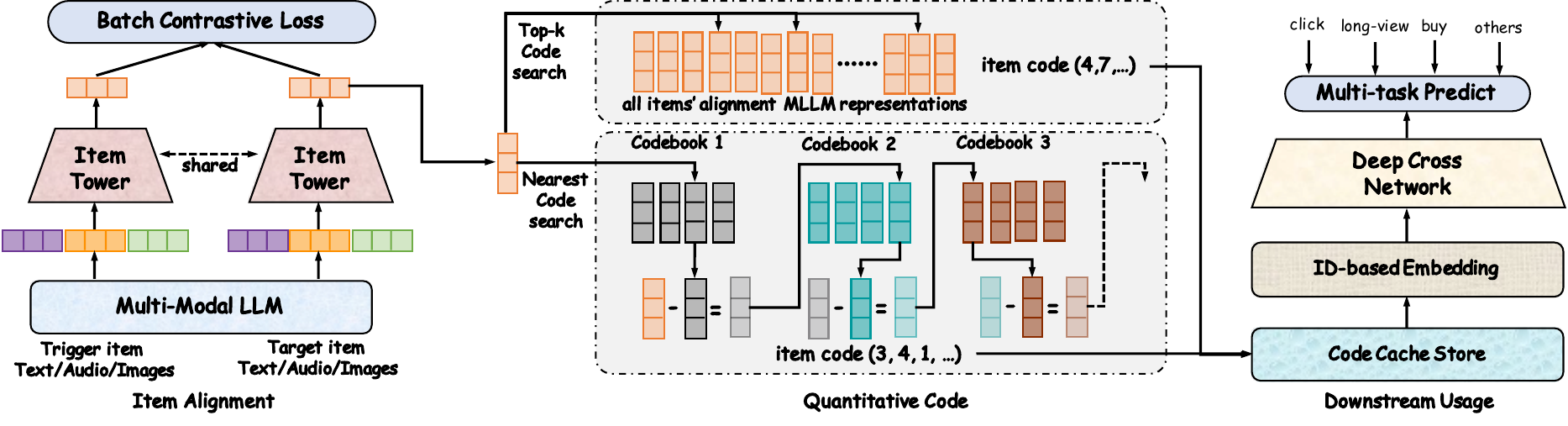}
  \caption{Sketches of QARM workflow, the left sub-figure gives the training framework of our item alignment mechanism, the middle sub-figure expresses our quantitative mechanism, and the right sub-figure indicates the learnable ID-based training.}
  \label{fig:qarm}
\end{figure*}
\subsection{Background of Industrial Recommender}
\label{modelfeature}
In industry, RecSys models are trained by the massive real-time user-item interaction data (i.e., the labels) and a large number of learnable discrete ID-based features (i.e., the model input)\cite{dlrm}.
To be specific, the model input features are always formed into four types:
\begin{itemize}[leftmargin=*,align=left]
    \item \textbf{ID-based features} to describe the context information, such as the User ID, Item ID, Scenario ID, is follow author, etc.
    \item \textbf{List-wise ID-based features} to describe user interests or item property, such as the users lastest clicked items, other similar item list with item candidate.
    \item \textbf{Bucketing ID-based features} to describe the statistics features, such as the numbers of purchases in the last month.
    \item \textbf{Multi-modal frozen representation features} to describe the items text/visual/audio information.
\end{itemize}
Equipped with above input features, to find the most related items from billion-scale item pool, the recommender chain always follow a two-stage cascading paradigm~\cite{youtube} to make a trade-off between performance and efficiency to respond the vast online requests: 
\begin{enumerate}[leftmargin=*,align=left]
\item Retrieval stage~\cite{yang2020large,liu2024kuaiformertransformerbasedretrievalkuaishou}: according user's past interested items, the Retrieval stage always uses simple models to search a small hundred items from total billion item set.
\item Ranking stage~\cite{crossmoment,lv2024marmunlockingfuturerecommendation}: based on generated item sets from retrieval stage, the ranking stage always utilizes complex model to play hundreds of times to estimate probabilities for hundreds of item candidates, and then select the best dozen items with the highest interaction probabilities.
\end{enumerate}
Generally, the Retrieval models follow two classic designs, the User2Item and Item2Item framework, while the Ranking model follows a multi-task learning paradigm, as shown in Figure~\ref{fig:background}.
In the following section, we express how our QARM enhances the retrieval and ranking stages.

\subsection{Item Alignment of QARM}
Particularly, the multi-modal representations are verified to contribute remarkable gains to serve as the user-side feature, item-side feature, and target item-aware user historical item list feature in our recommendation models.
Since the aforementioned serious representation unmatching issue, the multi-modal information is still limited.
To alleviate such problem, previous efforts always devise additional contrastive module to align with item ID and its multi-modal representations.
However, such contrastive loss is weak and easily over-fitting because of the ground-truth is not diverse enough, e.g., the item MLLM only has one ground-truth, its item ID embedding.

Instead of contrastive objective, to ensure that the multi-modal features are relevant to the user behavior decisions of specific businesses, we consider further fine-tuning such multi-modal representation with the the real downstream business interaction data before the representation input to downstream models.
To implement the above idea, we decide to build a pure multi-modal representation input only alignment model, and employ the existing retrieval models' knowledge to supervise it to reflect real business characteristics.
Specifically, we first generate high-quality item2item pairs dataset $\mathcal{D}$ in following ways:
\begin{itemize}[leftmargin=*,align=left]
    \item Based on User2Item retrieval model, for each user positive clicked target item, select the highest similar item in ID representation space as trigger item from historical lastest 50 his/her positive clicked item set.
    \item Based on Item2Item retrieval model, utilizing existing models learned stable item pairs with high similarity as data sources, e.g., export data from our Swing retrieval model.
\end{itemize}

On top of the high-quality item2item pairs dataset $\mathcal{D}$, we next train a item2item style alignment models with pure multi-modal representation.
For a random batch data $\mathcal{B}\in \mathcal{D}$, we have:
\begin{equation}
\begin{split}
\textbf{M}_{\texttt{trigger}} &= \texttt{MLLM}(\mathbf{T}^{\texttt{text}}_{\texttt{trigger}}, \mathbf{T}^{\texttt{audio}}_{\texttt{trigger}}, \mathbf{T}^{\texttt{image}}_{\texttt{trigger}}),\\
\textbf{M}_{\texttt{target}} &= \texttt{MLLM}(\mathbf{T}^{\texttt{text}}_{\texttt{target}}, \mathbf{T}^{\texttt{audio}}_{\texttt{target}}, \mathbf{T}^{\texttt{image}}_{\texttt{target}}),\\
\mathcal{L}_{\texttt{align}} = \texttt{Bat}&\texttt{ch-Contrastive}(\textbf{M}_{\texttt{trigger}}, \textbf{M}_{\texttt{target}}, \mathcal{B}),\\
\end{split}
\label{finetune}
\end{equation}
where the $\textbf{M}_{\texttt{trigger}}$/$\textbf{M}_{\texttt{target}}\in \mathbb{R}^{|\mathcal{B}|\times d}$ means the generated trigger/target item MLLM representation in batch manner ($d$ indicates representation dimension), the $\mathbf{T}^{\texttt{text}}_{\texttt{trigger}}$, $\mathbf{T}^{\texttt{audio}}_{\texttt{trigger}}$, $\mathbf{T}^{\texttt{image}}_{\texttt{trigger}}$/$\mathbf{T}^{\texttt{text}}_{\texttt{target}}$, $\mathbf{T}^{\texttt{audio}}_{\texttt{target}}$, $\mathbf{T}^{\texttt{image}}_{\texttt{target}}$ are raw input text, audio and image tokens of trigger/target item for MLLM, and the $\mathcal{L}_{\texttt{align}}$ is our QARM alignment training loss.
By optimizing the item alignment loss, the MLLM representations are encouraged to align with the downstream business knowledge, maximizing the representation consistency.

\subsection{Quantitative Code of QARM}
After obtaining the alignment multi-modal representation, the next stage is to apply the MLLM world knowledge to enhance the downstream models prediction accuracy.
However, comparing utilizing the pre-trained representation as a part of model input directly, the recommendation model is is actually more suitable for end-to-end training using ID style features.
Inspired by the code hashing and straight-through idea achieves great success in CV~\cite{vqgan} and DM~\cite{rajput2023recommender}, we also consider generating a series of quantitative code IDs to replace the MLLM representation.
Specifically, we design two heuristics simple-but-effective quantitative mechanism to transform the learned item alignment MLLM representation $\mathbf{M}\in \mathbb{R}^{|\mathcal{I}|\times d}$ by the Vector-Quantized (VQ)~\cite{van2017neural} and Residual-Quantized (RQ)~\cite{rqvae} codes, where the $\mathcal{I}$ denotes the item set.
\begin{itemize}[leftmargin=*,align=left]
    \item For the VQ code, as the most used quantitative technique, it first trains a large-scale codebook matrix and then utilizes the top-k nearest neighbor search to hash a representation. 
    In our QARM, since the pre-trained MLLM representations can already reflect complex items' correlation, thus we do not train a new codebook matrix but employ all the items' alignment representations as the codebook directly for the sake of simplicity:
    \begin{equation}
    \begin{split}
    \mathbf{V} = \mathbf{M},
    \end{split}
    \label{vqcodebook}
    \end{equation}
    where the $\mathbf{V}$ denotes the VQ codebook of QARM. According to it, we can quantization an arbitrary MLLM representation $\mathbf{m}\in \mathbf{M}$ as follows:
    \begin{equation}
    \begin{split}
    v_1, v_2, \dots, v_K = \texttt{TopKCode}(\mathbf{V}, \mathbf{m}, K),
    \end{split}
    \label{vqcodebook2}
    \end{equation}
    where the $\texttt{TopKCode}()$ aims to find the Topk similar representation index (i.e., code) from $\mathbf{V}$, $k$ is a hyper-parameter to control VQ codes number, and the $[v_1, v_2, \dots, v_k]$ is the VQ quantitative codes for representation $\mathbf{m}$.
    \item For the RQ code, instead of utilizing a larger codebook size to hash a representation, the RQ uses a fixed size of codebook to recursively quantize a representation in a coarse-to-fine manner.
    In practice, the RQ always trains $L$ levels codebooks with cascading relationships, and then searches the nearest neighbor index for each layer residual representation.
    In our QARM, we utilize the heuristics Kmeans algorithm to generate the codebook for each level:
    \begin{equation}
    \begin{split}
    \mathbf{R}^1 &= \texttt{Kmeans}(\mathbf{M}, N), \quad \mathbf{M}^1 = \mathbf{M}-\texttt{NearestRep}(\mathbf{M}, \mathbf{R}^1)\\
    \mathbf{R}^2 &= \texttt{Kmeans}(\mathbf{M}^1, N), \quad \mathbf{M}^2 = \mathbf{M}^1-\texttt{NearestRep}(\mathbf{M}^1, \mathbf{R}^2)\\
    &\dots,\quad \mathbf{R}^L = \texttt{Kmeans}(\mathbf{M}^{L-1}, N)\\
    \end{split}
    \label{rqcodebook}
    \end{equation}
    where the $\texttt{NearestRep}()$ denote the nearest representation search method in codebook, and the $[\mathbf{R}^1, \mathbf{R}^2, \dots, \mathbf{R}^L]$ are the trained codebook list of RQ.
    As a result, we can quantize an arbitrary item's MLLM representation $\mathbf{m}\in \mathbf{M}$ as follows:
    \begin{equation}
    \begin{split}
    r_1 &= \texttt{NearestCode}(\mathbf{R}^1, \mathbf{m}, 1), \quad \mathbf{m}^1 = \mathbf{m}-\mathbf{R}^1_{r_1}\\
    r_2 &= \texttt{NearestCode}(\mathbf{R}^2, \mathbf{m}^1, 1), \quad \mathbf{m}^2 = \mathbf{m}^1-\mathbf{R}^1_{r_2}\\
    &\dots,\quad r_L  = \texttt{NearestCode}(\mathbf{R}^L, \mathbf{m}^{L-1}, 1)\\
    \end{split}
    \label{rqcodebook}
    \end{equation}
    where the $[r_1, r_2, \dots, r_L]$ is the RQ codes for representation $\mathbf{m}$.
\end{itemize}

\begin{algorithm}[t]
\caption{Two type quantitative codebook pre-processing.}
\label{algo:codebook}
\begin{algorithmic}[1]
\REQUIRE $\mathbf{M}\in \mathbb{R}^{N\times d}$ denotes the fine-tuned multi-modal item representations, $N$ denotes item number, $d$ denotes dimension, $L$ denotes Codebook size for each layer.
\ENSURE the two type generated Codebook list.
\STATE \texttt{import faiss}
\STATE \texttt{VQCode = $\mathbf{M}$}
\STATE \texttt{RQCodeList = []}
\STATE \texttt{kmeans = faiss.Kmeans(d, L)}
\FOR {$i$ from $1 \rightarrow L$}
\STATE \texttt{kmeans.train(}$\mathbf{M}$\texttt{)}
\STATE \texttt{\_, I = kmeans.index.search(}$\mathbf{M}$\texttt{, 1)}
\STATE \texttt{I = I.reshape([-1])}
\STATE $\mathbf{O}$ \texttt{ = kmeans.centroids[I]}
\STATE $\mathbf{M}$ \texttt{=} $\mathbf{M}$ \texttt{-} $\mathbf{O}$
\STATE \texttt{RQCodeList.append(kmeans.centroids)}
\ENDFOR 
\STATE \texttt{Return VQCode, RQCodeList}
\end{algorithmic}
\end{algorithm}

The pseudo-code of our VQ and RQ codebook generation is shown in Algorithm~\ref{algo:codebook}, thereby any alignment MLLM representations could be transformed as two type codes $[v_1, v_2, \dots, v_K]$ and $[r_1, r_2, \dots, r_L]$, next we can save such codes to a cache store for end-to-end representation learning in recommendation model.

\subsection{Usage of QARM}

On top of the quantitative codes, we next devise several simple-but-effective ways to produce attributes to support our downstream recommendation model for end-to-end MLLM information training.
Generally, we implement the item-side feature, user-side feature for retrieval and ranking model, and the target item-aware feature for ranking model (as shown in the Figure~\ref{fig:background} and Section~\ref{modelfeature}):
\begin{itemize}[leftmargin=*,align=left]
    \item \textbf{Item-side feature}: Straightforwardly, we utilize the VQ code and RQ code as item ID feature, and then assign corresponding embedding spaces for these codes to lookup end-to-end learnable embeddings.
    \item \textbf{User-side feature}: To describe users' interests, we employ the quantitative codes of latest user's positive interacted items' sequence as a part of model input.
    \item \textbf{Target item-aware feature}: Instead of learning our code representation directly, we also apply the target item quantitative code to search several item sequences as target item-aware cross features.
    For example, according to our RQ code, we could generate the latest first one-code matching item sequence, the latest two-code matching item sequence, and so on.
\end{itemize}
The above features modeling methods are basically the same as some common works in the industry, and overall learning processes formed are as follows:
\begin{equation*}
\small
\begin{split}
\mathbf{Code}_i = \texttt{IDLook}&\texttt{Up}([v_1^i, \dots, v_K^i]\oplus[r_1^i, \dots, r_L^i]),\\
\texttt{ItemCodeRep} = &\texttt{ItemNet}(\mathbf{Code}_{\texttt{Target}}),\\
\texttt{UserCodeRep} = &\texttt{UserNet}([\mathbf{Code}_1, \dots, \mathbf{Code}_n]),\\
\texttt{CrossCodeRep} = \texttt{CrossNet}(&\texttt{ItemCodeRep}, [\mathbf{Code}_1^{\texttt{Search}}, \dots, \mathbf{Code}_n^{\texttt{Search}}]),\\
\hat{y}^\texttt{ctr}, \hat{y}^\texttt{lvtr}, ... = \texttt{MoE}([&\texttt{UserCodeRep}, \texttt{ItemCodeRep}, \\
&\texttt{CrossCodeRep}, \texttt{OtherFeaRep}])\\
\mathcal{L} = - \sum_{\texttt{xtr}}^{{\texttt{ctr}, \dots}} \big(y^{\texttt{xtr}}&\texttt{log}{(\hat{y}^{\texttt{xtr}})} + (1-y^{\texttt{xtr}})\texttt{log}{(1-\hat{y}^{\texttt{xtr}}})\big)
\end{split}
\label{modeling}
\end{equation*}
where the $\mathbf{Code}_i$ is the selected embedding of item $i$, \texttt{ItemCodeRep} means the item-side code features, \texttt{UserCodeRep} is the user-side code feature, the \texttt{CrossCodeRep} denote the target item-aware multi-modal feature, the \texttt{OtherFeaRep} denote the additional other user/item features' representations, the \texttt{MoE} is a multi-task prediction module, and the $\mathcal{L}$ is the training objective of our ranking model.

\begin{figure}[t]
  \centering
  \includegraphics[width=8cm,height=3cm]{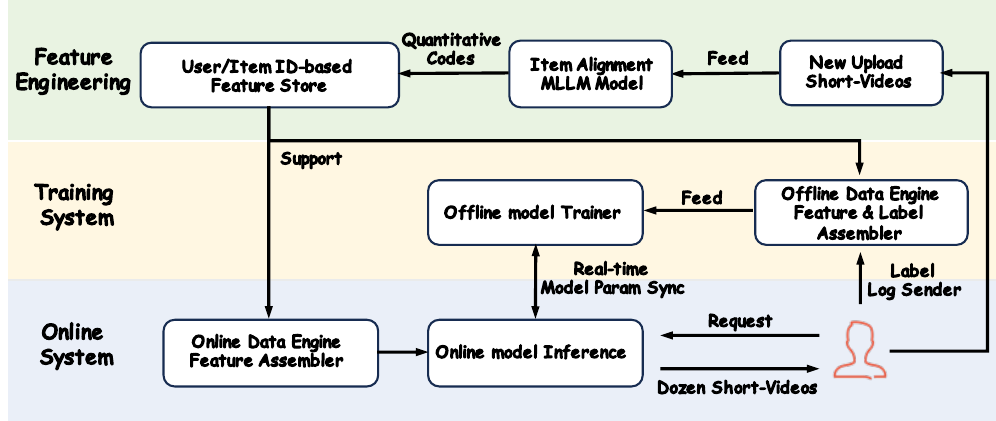}
  \caption{Overall system deployment pipeline.}
  \label{fig:deploy}
\end{figure}

\section{System Deployment}
As shown in Figure~\ref{fig:deploy}, our QARM is trained on Kuaishou's distributed training system, there are three basic blocks needed, the feature engineering, offline training system, and online serving system.
To be specific, our QARM is a part of the feature engineering block, which focuses on pre-processing the users' new uploaded short-video information as ID-based codes to support our offline training and online serving system.
To optimize such multi-modal code embedding, our offline data engine assembles the hundreds of billions user-item pairs feature and the real interaction label every day, to feed our model to update the model parameters.
Furthermore, the trained parameters are synchronized to the online inference model in real-time manner, to respond to the user's recommendation requests.

\begin{table*}[t!]
\centering
\caption{Offline results(\%) in term AUC, UAUC and GAUC on Advertising services at Kuaishou.}
\setlength{\tabcolsep}{18.5pt}{
\begin{tabular}{l|cccccc}
\toprule
\multirow{4}{*}{\makecell{QARM\\Model\\Variants}} 
& \multicolumn{6}{c}{Advertising}   \\ 
\cmidrule(r){2-7} & \multicolumn{3}{c}{CTR}  & \multicolumn{3}{c}{CVR}   \\ 
\cmidrule(r){2-4} \cmidrule(r){5-7} & AUC & UAUC & GAUC & AUC & UAUC & GAUC \\
\hline
Baseline Model & 90.85\% & 70.88\% & 72.43\% & 88.29\% & 69.01\% & 74.03\%  \\ 
+ IA Rep & +0.02\% & +0.05\% & -0.01\% & +0.03\% & +0.04\% & -0.02\%  \\
+ VQ Code  & +0.11\% & +0.20\% & +0.17\% & +0.09\% & +0.05\% & +0.11\% \\
+ RQ Code & +0.10\% & +0.15\% & +0.13\% & +0.07\% & +0.09\% & +0.10\%  \\
+ VQ \& RQ Code & +0.18\% & +0.29\% & +0.25\% & +0.17\% & +0.08\% & +0.12\%  \\
\bottomrule
\end{tabular}
}
\label{mainofflineadvertising}
\end{table*}

\begin{table*}[t!]
\centering
\caption{Offline results(\%) in term AUC, UAUC and GAUC on Shopping services at Kuaishou.}
\setlength{\tabcolsep}{9.5pt}{
\begin{tabular}{l|ccccccccc}
\toprule
\multirow{4}{*}{\makecell{QARM\\Model\\Variants}} 
& \multicolumn{9}{c}{Shopping}   \\ 
\cmidrule(r){2-10} & \multicolumn{3}{c}{CTR}  & \multicolumn{3}{c}{CVR} & \multicolumn{3}{c}{CTCVR}  \\ 
\cmidrule(r){2-4} \cmidrule(r){5-7} \cmidrule(r){8-10} & AUC & UAUC & GAUC & AUC & UAUC & GAUC & AUC & UAUC & GAUC\\
\hline
Baseline Model & 79.31\% & 66.48\% & 68.05\% & 86.79\% & 69.13\% & 69.82\% & 87.92\% & 70.70\% & 71.41\%  \\ 
+ IA Rep & -0.03\% & +0.02\% & -0.01\% & -0.02\% & +0.02\% & +0.03\% & -0.07\% & +0.13\% &+0.020\%  \\
+ VQ Code  & +0.09\% & +0.20\% & +0.21\% & +0.11\% & +0.38\% & +0.38\% & +0.19\% & +0.49\% & +0.52\%\\
+ RQ Code  & +0.12\% & +0.33\% & +0.26\% & +0.14\% & +0.32\% & +0.34\% & +0.18\% & +0.50\% & +0.40\%\\
+ VQ \& RQ Code & +0.23\% & +0.56\% & +0.50\% & +0.26\% & +0.77\% & +0.77\% & +0.35\% & +1.00\% & +0.92\% \\
\bottomrule
\end{tabular}
}
\label{mainofflineshop}
\end{table*}

\section{Experiments}
In this section, we conduct detailed offline/online experiments and detailed ablation studies at Kuaishou's Shopping and Advertising services, to validate our QARM effectiveness.

\subsection{Evaluation Technique}
As a common evaluation protocol in industry~\cite{dien,twin}, to verify how much benefit that our QARM can contribute to our system, we equip it to our baseline models at Shopping and Advertising services.
It is worth noting that the two strong baseline models are huge models and have been already incorporated the cached MLLM representation in Figure~\ref{fig:intro}.
For the evaluation, we use three wide-used ranking metrics to measure ranking model prediction performance: the AUC, UAUC and GAUC.
The AUC metric reflects the general probability that the score of a positive user-item pair is higher than the score of a negative user-item pair. 
The UAUC metric estimates the average AUC value across different users. 
Moreover, the weighted version of UAUC, GAUC, incorporates different user interaction ratios to provide more precise evaluation.
They are formed as follows:
\begin{equation*}
\begin{split}
\texttt{UAUC} = \sum_{\texttt{u}}^{|\mathcal{U}|}\frac{1}{|\mathcal{U}|} \texttt{AUC}_{\texttt{u}},\quad
\texttt{GAUC} = \sum_{\texttt{u}}^{|\mathcal{U}|}\frac{\texttt{sample}_{\texttt{u}}}{\texttt{all sample}} \texttt{AUC}_{\texttt{u}},
\end{split}
\label{modeling}
\end{equation*}
where $\texttt{sample}_{\texttt{u}}$ denote the all user-item pair belongs to the user $\texttt{u}$.

\begin{table}[ht]
\centering
\caption{Online A/B testing results of Advertising services.}
\setlength{\tabcolsep}{3.5pt}{

\begin{tabular}{c|c|ccc}
\toprule
\multirow{2}{*}{Scenarios}  & \multirow{2}{*}{\makecell{Item\\Group}}                  & \multicolumn{3}{c}{Advertising Metrics}                                                                             \\
\cmidrule(r){3-5}  
    & & Exposure &  Cost   & Revenue \\
\midrule
\multirow{2}{*}{\makecell{Advertising\#1}}  & Cold-start & +5.324\% & +7.676\% &  +9.704\% \\
 & Others & +1.909\% & +3.383\% & +3.147\% \\
\midrule
\multirow{2}{*}{\makecell{Advertising\#2}} & Cold-start & +9.876\% &  +10.570\% & +9.555\% \\
& Others &+0.911\% & +1.738\%& +1.950\% \\
\bottomrule
\end{tabular}
}
\label{mainonlineadv}
\end{table}

\begin{table}[ht]
\centering
\caption{Online A/B testing results of Shopping services.}
\setlength{\tabcolsep}{3.5pt}{

\begin{tabular}{c|ccccc}
\toprule
\multirow{2}{*}{Scenarios}                    & \multicolumn{5}{c}{Shopping Metrics}                                                                                  \\
\cmidrule(r){2-6}  
    &Orders & GMV &  CTR   & CVR & GPM \\
\midrule
Shopping\#1 &+1.396\% &+2.296\% &+0.478\% &+0.903\% &+0.553\% \\
Shopping\#2 &+0.716\% &+1.568\% &+1.007\% &-0.000\% &+3.110\% \\
\bottomrule
\end{tabular}
}
\label{mainonlineshopping}
\end{table}

\begin{figure}[t]
  \centering
  \includegraphics[width=8cm,height=6cm]{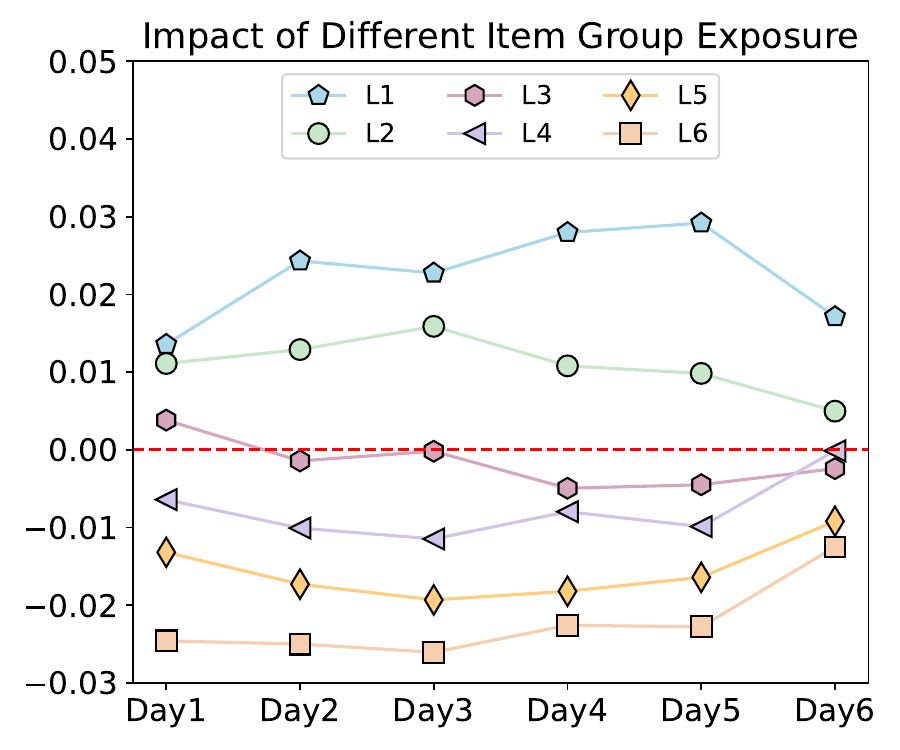}
  \caption{The exposure trend across different item groups.}
  \label{iaablimpression}
  \vspace{-5mm}
\end{figure}

\begin{figure*}[t]
  \centering
  \includegraphics[width=16cm,height=6cm]{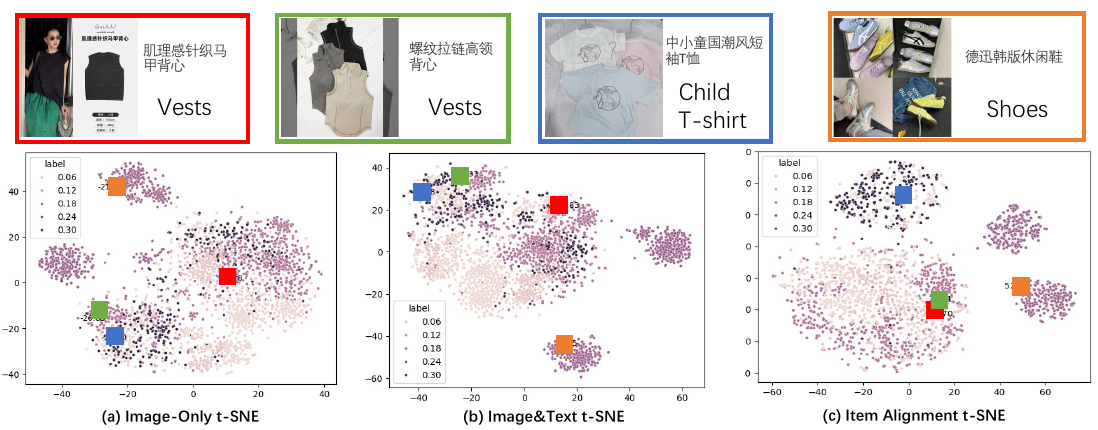}
  \caption{Visualization of the item representations from the pre-trained MLLM(a)(b), and item alignment fine-tuned MLLM(c).}
  \label{iaabltsne}
\end{figure*}

\subsection{Offline Performance}
Table~\ref{mainofflineshop} and Table~\ref{mainofflineadvertising} report the shopping and advertising services‘ offline results in terms of the CTR, CVR (we set the VQ code length is $K=25$, RQ code length is $L=6$, dimension $d=64$).
From them, we have several observations:
(1) Compared with the different Baseline Model, directly incorporating the item alignment MLLM representation could further enhance the models prediction accuracy in the Advertising services, which indicates that alleviates the representation unmatching issue to encourage MLLM representation to align with the real bussiness interaction distribution knowledge is helpful.
(2) Compared with `+ IA Rep' model variant, utilizing the VQ code or RQ code to represent the items' MLLM representation are shown more significant improvements, which indicates that overcoming the representation unlearning issue and assigning learnable embedding for end-to-end training is vital for recommender model convergency.
(3) Compared with the `+ VQ code' or `+ RQ code' model variants, incorporating the VQ and RQ codes at same time could further enhance model performance, we think the reason might be that the two codes could reflect different MLLM knowledge in our designing.
Indeed, the VQ code aims to utilize the TopK similar item neighbors to represent target item information, while the RQ code focuses on encoding the entire MLLM representation into a hierarchical residual path to represent target item information.

\begin{table}[t!]
\centering
\caption{Offline effect of QARM across different item groups.}
\setlength{\tabcolsep}{2pt}{
\begin{tabular}{l|ccccccc}
\toprule
\multirow{3}{*}{\makecell{Metric}}
& \multicolumn{7}{c}{Shopping Item Groups}  \\ 
\cmidrule(r){2-8}   & L1 & L2 & L3 & L4 & L5 & L6 & ALL\\
\hline
CTR-AUC & +0.26\ & +0.25\% &+0.23\%  &+0.22\% &+0.24\% &+0.21\% & +0.23\% \\
\bottomrule
\end{tabular}
}
\label{iaalb}
\end{table}

\subsection{Online A/B Test}
To quantify the certain contribution our QARM could make to our online services, we launched the variant `+ VQ \& RQ Code' to online A/B test system to respond real user Shopping and Advertising requests.
In practice, different services have their core revenue metrics, e.g., the ePCM of advertising and the GMV of shopping.
Table~\ref{mainonlineadv} and Table~\ref{mainonlineshopping} report our online results of Advertising and Shopping services individually, where the `\#1' and `\#2' denote different application scenarios of corresponding services, e.g., short-video scenarios or Mall scenario.
According to them, we can find that our QARM achieves a large improvement at Revenue +9.704\%/+3.147\% and +9.555\%/+1.950\% on the cold-start item group and another group in advertising,
and GMV+2.296\%/1.568\% in online-shopping, which indicates understand the multi-modal item semantic signal could contribute to our system significantly.

\subsection{Case Study of QARM}
This section explores the impact of the multi-modal information on our system content distribution.
Intuitively, the multi-modal information is friendly to long-tail items since multi-modal information is not sensitive to whether the items are popular or not.
Therefore, the improvement in the performance of long-tail items can to some extent prove that our QARM is a reasonable way to introduce multi-modal information for RecSys.
To valid our QARM effectiveness in long-tail items, we divide items into six distinct groups, labeled from L1 to L6, based on the frequency of their purchases. L1 represents the least frequently purchased items, with each subsequent label indicating higher purchase frequencies.
Firstly, we conduct the offline evaluation in the Table~\ref{iaalb}, from that we could observe a phenomenon: in terms of the CTR-AUC, the most important offline metric, the long-tail items show the highest prediction improvement and slowly decays according to the purchased times.
Secondly, we also conduct the online evaluation in the Figure~\ref{iaablimpression}, which describes the item exposure times improvement per day.
According to Figure~\ref{iaablimpression}, compared to the baseline model, the L1, and L2 items are enhanced to successfully recommend more times, while the more popular L4, L5, and L6 items are less frequently.
Such two phenomenons indicate our QARM is a trustworthy way to inject the multi-modal information for RecSys, not only bringing a better experience to users but also building a fair environment for new-uploaded items.

\begin{table}[t]
\centering
\caption{Online effect of QARM across different item groups.}
\setlength{\tabcolsep}{3.5pt}{

\begin{tabular}{c|ccccc}
\toprule
\multirow{3}{*}{\makecell{Item\\Group}}               & \multicolumn{5}{c}{Shopping Metrics}                                                                                  \\
\cmidrule(r){2-6}  
    &Orders & GMV &  CTR   & CVR & GPM \\
\midrule
ALL &+0.716\% &+1.568\% &+1.007\% &-0.000\% &+3.110\% \\
\hline
L1 & 5.381\% & 8.035\% & 0.035\% &3.351\% &8.035\% \\
L2 &1.971\% &2.617\%& 0.312\%& 0.239\%& 2.617\% \\
L3 &0.249\% &1.207\% &1.015\% &-0.647\% &1.207\% \\
L4 &0.967\% &3.054\% &1.156\% &0.437\% &3.054\% \\
L5 &-0.407\% &1.317\% &0.909\% &0.069\% &1.317\% \\
L6 &1.231\% &3.984\% &1.691\% &1.269\% &3.984\% \\
\bottomrule
\end{tabular}
}
\label{iaalb2}
\vspace{-5mm}
\end{table}

Moreover, we further conduct an experiment to visualize the impact of item alignment MLLM representations.
Figure~\ref{iaabltsne} gives three different t-SNE results under the mall scenario of our shopping service: (1) original image-only t-SNE, (2) original image-text t-SNE, and (3) item alignment t-SNE.
Besides, we randomly select four items to show our item alignment effect, the two vests, one child T-shirt and a shoes.
According to Figure~\ref{iaabltsne}, we can find that the three methods could identify the shoes and clothes satisfactorily.
Nevertheless, the original MLLM representations failed to measure the two vests correctly, while the blue-colored child T-shirt is more closely with green-colored vests since they have similar image styles.
In our item alignment version, the two different vests are mapped to the closest location, which reveals our item alignment representation could reflect the business characteristics successfully, maximizing the representation consistency.

\subsection{Parameter Analysis of Quantitative Code}
Specifically, we set the VQ code length $K=25$, RQ code length $L=6$, and dimension $d=64$ by default, to investigate the robustness of our QARM regarding the hyper-parameter factors of quantitative codes, we conduct experiments on the VQ code length $K$ and the code embedding dimension $d$.
For the VQ code length $K$, we implement 4 variants on the shopping scenario and shown the results in Table~\ref{ablvqcode}.
According to it, we could observe that the VQ code length contributes in our model provides a relatively stable improvement to our QARM.
For the code embedding dimension $d$, we implement 2 variants on the advertising scenario and shown the results in Table~\ref{ablrqcode}.
According to it, we could observe that increasing the code dimension is also an effective way to enhance model performance.
In summary, the experiment results demonstrate that learning a down-streaming task adaptive representation is vital for recommendation to understand the multi-modal information.

\begin{table}[t!]
\centering
\caption{Ablation study of Quantitative VQ code length.}
\setlength{\tabcolsep}{13pt}{
\begin{tabular}{l|ccc}
\toprule
\multirow{4}{*}{\makecell{QARM\\Variants}} 
& \multicolumn{3}{c}{Shopping} \\ 
\cmidrule(r){2-4}  & \multicolumn{3}{c}{CTR}  \\ 
\cmidrule(r){2-4}   & AUC & UAUC & GAUC \\
\hline
+ VQ Code ($K$=5) & +0.02\%	 & +0.09\% &+0.08\% \\
+ VQ Code ($K$=10) & +0.03\% &+0.10\% &+0.10\% \\
+ VQ Code ($K$=15) & +0.04\% &+0.12\%	&+0.13\% \\
+ VQ Code ($K$=20) & +0.06\% &+0.16\%	&+0.16\% \\
\hline
+ VQ Code ($K$=25) & +0.09\% & +0.20\% & +0.21\% \\
\bottomrule
\end{tabular}
}
\label{ablvqcode}
\vspace{-5mm}
\end{table}

\section{Related Works}

\textit{Contrasitve learning based multi-modal information fusion}
Initial approaches involve utilizing "off-the-shelf" multimodal representations, either as fixed features or in conjunction with structural relationships, within the recommendation framework. For example, VBPR\cite{vbpr} enhances Matrix Factorization by incorporating visual features through a linear transform kernel, which is then concatenated with ID embedding. LATTICE\cite{lattice} proposes to build visual and textual affinity graphs using respective embeddings and provide multimodal item-item relationships to the collaborative filtering model. BM3\cite{bm3} leverages self-supervised learning to align both inter-modality and intra-modality representations within the collaborative filtering task.

To further improve modality-driven recommendations, multimodal pre-training is crucial. DVBPR\cite{dvbpr} extends VBPR by jointly training CNN visual encoder with the Matrix Factorization task. AlignRec\cite{alignrec} proposes pre-training the visual-text alignment task using a mask-then-predict strategy, ~~with BEiT3\cite{beit-3} as the backbone,~~ then align with ID representation in collaborative filtering task based on fixed multimodal representations. Sheng et al\cite{simtier} proposes semantic-aware contrastive learning in the pre-training phase, utilizing user's search query and subsequent purchase item to form positive sample pairs, then extracting features using SimTier and MAKE based on fixed multimodal representations. In our methods, we step further by aligning multimodal representation with downstream business-specific item-item relationships in pre-training, and leveraging quantitative code mechanisms for end-to-end training in recommendation model.

\textit{Quantitative multi-modal representation for recommendation}
Discrete quantization representations facilitate the precise approximation of a vector by decomposing it into multiple discrete code representations. This approach has found extensive applications across various domains~\cite{wright2010sparse, babenko2014additive, li2021trq, li2017performance}. Product Quantization (PQ) is used to compress high dimensional vectors by dividing them into subvectors and independently quantizing each subvector\cite{gray1984vector}. Residual Quantization (RQ) is a generalization of Product Quantization which focuses on quantizing the residuals left after the previous quantization~\cite{martinez2014stacked, ferdowsi2017regularized}. This method aims to improve the accuracy of the quantized representation.

In recommender systems, content discrete representations are also widely used ~\cite{hou2023learning, singh2024better, rajput2023recommender}. Using content discrete representations also known as semantic IDs, facilitates collisions between semantically related items,  thereby enhancing generalization in recommender models. TIGER\cite{rajput2023recommender} employed RQ-VAE~\cite{rqvae} to discretize the content embedding presentations of an item. Subsequently, an autoregressive model is utilized to predict the semantic ID of the next item that user interested. Singh et al.\cite{singh2024better} illustrated that hierarchical Semantic IDs can replace item IDs in ranking models and achieve superior generalization outcomes.

\begin{table}[t!]
\centering
\caption{Ablation study of Quantitative code dimension.}
\setlength{\tabcolsep}{10pt}{
\begin{tabular}{l|ccc}
\toprule
\multirow{4}{*}{\makecell{QARM\\Variants}} 
& \multicolumn{3}{c}{Advertising}   \\ 
\cmidrule(r){2-4} & \multicolumn{3}{c}{CTR} \\ 
\cmidrule(r){2-4} & AUC & UAUC & GAUC \\
\hline
+ VQ\&RQ Code($d$=16) & +0.11\% & +0.13\% & +0.11\% \\
+ VQ\&RQ Code($d$=32) & +0.16\% & +0.20\% & +0.17\% \\
\hline
+ VQ\&RQ Code($d$=64) & +0.18\% & +0.29\% & +0.25\% \\
\bottomrule
\end{tabular}
}
\label{ablrqcode}
\vspace{-5mm}
\end{table}

\section{Conclusion}
In this paper, we present a novel approach to injecting multi-modal information to recommendation model, QARM.
Different from the common deployment paradigm which utilizes fixed unlearnable pre-trained MLLM representation, our QARM utilizes the quantitative code of down-streaming task alignment fine-tuned MLLM representations to reach an end-to-end MLLM information training.
Specifically, in the item alignment mechanism, we first mine a group of high-quality down-streaming task item-item pairs, and then utilize them to guide the MLLM fine-tuning.
For the quantitative code mechanism, we devise two heuristic VQ and RQ code methods to quantify those representations to construct the user-side, item-side and target item-aware features to achieve end-to-end optimization for better convergence.
Empirically experimental results on Kuaishou's advertising and shopping scenarios demonstrate the effectiveness of QARM on multi-modal information fusion.
Besides, detailed analyses show that our QARM is more friendly for cold-start and long-tailed items, which is expected for multi-modal information usage.
Our QARM has been deployed at Kuaishou to support various services from 2024 March, serving 400 Million users every day.

\balance
\bibliographystyle{ACM-Reference-Format}
\bibliography{sample-base-extend.bib}


\begin{thebibliography}{36}


\ifx \showCODEN    \undefined \def \showCODEN     #1{\unskip}     \fi
\ifx \showDOI      \undefined \def \showDOI       #1{#1}\fi
\ifx \showISBNx    \undefined \def \showISBNx     #1{\unskip}     \fi
\ifx \showISBNxiii \undefined \def \showISBNxiii  #1{\unskip}     \fi
\ifx \showISSN     \undefined \def \showISSN      #1{\unskip}     \fi
\ifx \showLCCN     \undefined \def \showLCCN      #1{\unskip}     \fi
\ifx \shownote     \undefined \def \shownote      #1{#1}          \fi
\ifx \showarticletitle \undefined \def \showarticletitle #1{#1}   \fi
\ifx \showURL      \undefined \def \showURL       {\relax}        \fi
\providecommand\bibfield[2]{#2}
\providecommand\bibinfo[2]{#2}
\providecommand\natexlab[1]{#1}
\providecommand\showeprint[2][]{arXiv:#2}

\bibitem[Babenko and Lempitsky(2014)]%
        {babenko2014additive}
\bibfield{author}{\bibinfo{person}{Artem Babenko} {and} \bibinfo{person}{Victor Lempitsky}.} \bibinfo{year}{2014}\natexlab{}.
\newblock \showarticletitle{Additive quantization for extreme vector compression}. In \bibinfo{booktitle}{\emph{IEEE/CVF Computer Vision and Pattern Recognition Conference (CVPR)}}.
\newblock


\bibitem[Brown(2020)]%
        {gpt2}
\bibfield{author}{\bibinfo{person}{Tom~B Brown}.} \bibinfo{year}{2020}\natexlab{}.
\newblock \showarticletitle{Language models are few-shot learners}.
\newblock \bibinfo{journal}{\emph{arXiv}} (\bibinfo{year}{2020}).
\newblock


\bibitem[Cao et~al\mbox{.}(2024)]%
        {crossmoment}
\bibfield{author}{\bibinfo{person}{Jiangxia Cao}, \bibinfo{person}{Shen Wang}, \bibinfo{person}{Yue Li}, \bibinfo{person}{Shenghui Wang}, \bibinfo{person}{Jian Tang}, \bibinfo{person}{Shiyao Wang}, \bibinfo{person}{Shuang Yang}, \bibinfo{person}{Zhaojie Liu}, {and} \bibinfo{person}{Guorui Zhou}.} \bibinfo{year}{2024}\natexlab{}.
\newblock \showarticletitle{Moment\&Cross: Next-Generation Real-Time Cross-Domain CTR Prediction for Live-Streaming Recommendation at Kuaishou}.
\newblock \bibinfo{journal}{\emph{arXiv}} (\bibinfo{year}{2024}).
\newblock


\bibitem[Chang et~al\mbox{.}(2023)]%
        {twin}
\bibfield{author}{\bibinfo{person}{Jianxin Chang}, \bibinfo{person}{Chenbin Zhang}, \bibinfo{person}{Zhiyi Fu}, \bibinfo{person}{Xiaoxue Zang}, \bibinfo{person}{Lin Guan}, \bibinfo{person}{Jing Lu}, \bibinfo{person}{Yiqun Hui}, \bibinfo{person}{Dewei Leng}, \bibinfo{person}{Yanan Niu}, \bibinfo{person}{Yang Song}, {and} \bibinfo{person}{Kun Gai}.} \bibinfo{year}{2023}\natexlab{}.
\newblock \showarticletitle{TWIN: TWo-stage Interest Network for Lifelong User Behavior Modeling in CTR Prediction at Kuaishou}. In \bibinfo{booktitle}{\emph{ACM SIGKDD Conference on Knowledge Discovery and Data Mining (KDD)}}.
\newblock


\bibitem[Covington et~al\mbox{.}(2016)]%
        {youtube}
\bibfield{author}{\bibinfo{person}{Paul Covington}, \bibinfo{person}{Jay Adams}, {and} \bibinfo{person}{Emre Sargin}.} \bibinfo{year}{2016}\natexlab{}.
\newblock \showarticletitle{Deep Neural Networks for YouTube Recommendations}. In \bibinfo{booktitle}{\emph{ACM Conference on Recommender Systems (RecSys)}}.
\newblock


\bibitem[Esser et~al\mbox{.}(2021)]%
        {vqgan}
\bibfield{author}{\bibinfo{person}{Patrick Esser}, \bibinfo{person}{Robin Rombach}, {and} \bibinfo{person}{Bjorn Ommer}.} \bibinfo{year}{2021}\natexlab{}.
\newblock \showarticletitle{Taming transformers for high-resolution image synthesis}. In \bibinfo{booktitle}{\emph{IEEE/CVF Computer Vision and Pattern Recognition Conference (CVPR)}}.
\newblock


\bibitem[Ferdowsi et~al\mbox{.}(2017)]%
        {ferdowsi2017regularized}
\bibfield{author}{\bibinfo{person}{Sohrab Ferdowsi}, \bibinfo{person}{Slava Voloshynovskiy}, {and} \bibinfo{person}{Dimche Kostadinov}.} \bibinfo{year}{2017}\natexlab{}.
\newblock \showarticletitle{Regularized Residual Quantization: a multi-layer sparse dictionary learning approach}.
\newblock \bibinfo{journal}{\emph{arXiv}} (\bibinfo{year}{2017}).
\newblock


\bibitem[Gray(1984)]%
        {gray1984vector}
\bibfield{author}{\bibinfo{person}{Robert Gray}.} \bibinfo{year}{1984}\natexlab{}.
\newblock \showarticletitle{Vector quantization}.
\newblock \bibinfo{journal}{\emph{IEEE Assp Magazine}} (\bibinfo{year}{1984}).
\newblock


\bibitem[He and McAuley(0116)]%
        {vbpr}
\bibfield{author}{\bibinfo{person}{Ruining He} {and} \bibinfo{person}{Julian McAuley}.} \bibinfo{year}{20116}\natexlab{}.
\newblock \showarticletitle{VBPR: visual Bayesian Personalized Ranking from implicit feedback}. In \bibinfo{booktitle}{\emph{AAAI Conference on Artificial Intelligence (AAAI)}}.
\newblock


\bibitem[Hou et~al\mbox{.}(2023)]%
        {hou2023learning}
\bibfield{author}{\bibinfo{person}{Yupeng Hou}, \bibinfo{person}{Zhankui He}, \bibinfo{person}{Julian McAuley}, {and} \bibinfo{person}{Wayne~Xin Zhao}.} \bibinfo{year}{2023}\natexlab{}.
\newblock \showarticletitle{Learning vector-quantized item representation for transferable sequential recommenders}. In \bibinfo{booktitle}{\emph{International World Wide Web Conference (WWW)}}.
\newblock


\bibitem[Kang et~al\mbox{.}(2017)]%
        {dvbpr}
\bibfield{author}{\bibinfo{person}{Wang-Cheng Kang}, \bibinfo{person}{Chen Fang}, \bibinfo{person}{Zhaowen Wang}, {and} \bibinfo{person}{Julian McAuley}.} \bibinfo{year}{2017}\natexlab{}.
\newblock \showarticletitle{Visually-Aware Fashion Recommendation and Design with Generative Image Models}. In \bibinfo{booktitle}{\emph{IEEE International Conference on Data Mining (ICDM)}}.
\newblock


\bibitem[Lee et~al\mbox{.}(2022)]%
        {rqvae}
\bibfield{author}{\bibinfo{person}{Doyup Lee}, \bibinfo{person}{Chiheon Kim}, \bibinfo{person}{Saehoon Kim}, \bibinfo{person}{Minsu Cho}, {and} \bibinfo{person}{Wook-Shin Han}.} \bibinfo{year}{2022}\natexlab{}.
\newblock \showarticletitle{Autoregressive image generation using residual quantization}. In \bibinfo{booktitle}{\emph{IEEE/CVF Computer Vision and Pattern Recognition Conference (CVPR)}}.
\newblock


\bibitem[Leskovec et~al\mbox{.}(2020)]%
        {leskovec2020mining}
\bibfield{author}{\bibinfo{person}{Jure Leskovec}, \bibinfo{person}{Anand Rajaraman}, {and} \bibinfo{person}{Jeffrey~David Ullman}.} \bibinfo{year}{2020}\natexlab{}.
\newblock \bibinfo{booktitle}{\emph{Mining of massive data sets}}.
\newblock \bibinfo{publisher}{Cambridge university press}.
\newblock


\bibitem[Li et~al\mbox{.}(2023)]%
        {blip2}
\bibfield{author}{\bibinfo{person}{Junnan Li}, \bibinfo{person}{Dongxu Li}, \bibinfo{person}{Silvio Savarese}, {and} \bibinfo{person}{Steven Hoi}.} \bibinfo{year}{2023}\natexlab{}.
\newblock \showarticletitle{BLIP-2: bootstrapping language-image pre-training with frozen image encoders and large language models}. In \bibinfo{booktitle}{\emph{International Conference on Machine Learning (ICML)}}.
\newblock


\bibitem[Li et~al\mbox{.}(2021)]%
        {li2021trq}
\bibfield{author}{\bibinfo{person}{Yue Li}, \bibinfo{person}{Wenrui Ding}, \bibinfo{person}{Chunlei Liu}, \bibinfo{person}{Baochang Zhang}, {and} \bibinfo{person}{Guodong Guo}.} \bibinfo{year}{2021}\natexlab{}.
\newblock \showarticletitle{Trq: Ternary neural networks with residual quantization}. In \bibinfo{booktitle}{\emph{AAAI Conference on Artificial Intelligence (AAAI)}}.
\newblock


\bibitem[Li et~al\mbox{.}(2017)]%
        {li2017performance}
\bibfield{author}{\bibinfo{person}{Zefan Li}, \bibinfo{person}{Bingbing Ni}, \bibinfo{person}{Wenjun Zhang}, \bibinfo{person}{Xiaokang Yang}, {and} \bibinfo{person}{Wen Gao}.} \bibinfo{year}{2017}\natexlab{}.
\newblock \showarticletitle{Performance guaranteed network acceleration via high-order residual quantization}. In \bibinfo{booktitle}{\emph{International Conference on Computer Vision (ICCV)}}.
\newblock


\bibitem[Liu et~al\mbox{.}(2024a)]%
        {liu2024kuaiformertransformerbasedretrievalkuaishou}
\bibfield{author}{\bibinfo{person}{Chi Liu}, \bibinfo{person}{Jiangxia Cao}, \bibinfo{person}{Rui Huang}, \bibinfo{person}{Kai Zheng}, \bibinfo{person}{Qiang Luo}, \bibinfo{person}{Kun Gai}, {and} \bibinfo{person}{Guorui Zhou}.} \bibinfo{year}{2024}\natexlab{a}.
\newblock \showarticletitle{KuaiFormer: Transformer-Based Retrieval at Kuaishou}.
\newblock \bibinfo{journal}{\emph{arXiv}} (\bibinfo{year}{2024}).
\newblock


\bibitem[Liu et~al\mbox{.}(2024b)]%
        {alignrec}
\bibfield{author}{\bibinfo{person}{Yifan Liu}, \bibinfo{person}{Kangning Zhang}, \bibinfo{person}{Xiangyuan Ren}, \bibinfo{person}{Yanhua Huang}, \bibinfo{person}{Jiarui Jin}, \bibinfo{person}{Yingjie Qin}, \bibinfo{person}{Ruilong Su}, \bibinfo{person}{Ruiwen Xu}, \bibinfo{person}{Yong Yu}, {and} \bibinfo{person}{Weinan Zhang}.} \bibinfo{year}{2024}\natexlab{b}.
\newblock \showarticletitle{AlignRec: Aligning and Training in Multimodal Recommendations}. In \bibinfo{booktitle}{\emph{ACM International Conference on Information and Knowledge Management (CIKM)}}.
\newblock


\bibitem[Lv et~al\mbox{.}(2024)]%
        {lv2024marmunlockingfuturerecommendation}
\bibfield{author}{\bibinfo{person}{Xiao Lv}, \bibinfo{person}{Jiangxia Cao}, \bibinfo{person}{Shijie Guan}, \bibinfo{person}{Xiaoyou Zhou}, \bibinfo{person}{Zhiguang Qi}, \bibinfo{person}{Yaqiang Zang}, \bibinfo{person}{Ming Li}, \bibinfo{person}{Ben Wang}, \bibinfo{person}{Kun Gai}, {and} \bibinfo{person}{Guorui Zhou}.} \bibinfo{year}{2024}\natexlab{}.
\newblock \showarticletitle{MARM: Unlocking the Future of Recommendation Systems through Memory Augmentation and Scalable Complexity}.
\newblock \bibinfo{journal}{\emph{arXiv}} (\bibinfo{year}{2024}).
\newblock


\bibitem[Martinez et~al\mbox{.}(2014)]%
        {martinez2014stacked}
\bibfield{author}{\bibinfo{person}{Julieta Martinez}, \bibinfo{person}{Holger~H Hoos}, {and} \bibinfo{person}{James~J Little}.} \bibinfo{year}{2014}\natexlab{}.
\newblock \showarticletitle{Stacked quantizers for compositional vector compression}.
\newblock \bibinfo{journal}{\emph{arXiv}} (\bibinfo{year}{2014}).
\newblock


\bibitem[Naumov et~al\mbox{.}(2019)]%
        {dlrm}
\bibfield{author}{\bibinfo{person}{Maxim Naumov}, \bibinfo{person}{Dheevatsa Mudigere}, \bibinfo{person}{Hao-Jun~Michael Shi}, \bibinfo{person}{Jianyu Huang}, \bibinfo{person}{Narayanan Sundaraman}, \bibinfo{person}{Jongsoo Park}, \bibinfo{person}{Xiaodong Wang}, \bibinfo{person}{Udit Gupta}, \bibinfo{person}{Carole-Jean Wu}, \bibinfo{person}{Alisson~G. Azzolini}, \bibinfo{person}{Dmytro Dzhulgakov}, \bibinfo{person}{Andrey Mallevich}, \bibinfo{person}{Ilia Cherniavskii}, \bibinfo{person}{Yinghai Lu}, \bibinfo{person}{Raghuraman Krishnamoorthi}, \bibinfo{person}{Ansha Yu}, \bibinfo{person}{Volodymyr Kondratenko}, \bibinfo{person}{Stephanie Pereira}, \bibinfo{person}{Xianjie Chen}, \bibinfo{person}{Wenlin Chen}, \bibinfo{person}{Vijay Rao}, \bibinfo{person}{Bill Jia}, \bibinfo{person}{Liang Xiong}, {and} \bibinfo{person}{Misha Smelyanskiy}.} \bibinfo{year}{2019}\natexlab{}.
\newblock \showarticletitle{Deep Learning Recommendation Model for Personalization and Recommendation Systems}.
\newblock \bibinfo{journal}{\emph{arXiv}}.
\newblock


\bibitem[Radford et~al\mbox{.}(2021)]%
        {clip}
\bibfield{author}{\bibinfo{person}{Alec Radford}, \bibinfo{person}{Jong~Wook Kim}, \bibinfo{person}{Chris Hallacy}, \bibinfo{person}{Aditya Ramesh}, \bibinfo{person}{Gabriel Goh}, \bibinfo{person}{Sandhini Agarwal}, \bibinfo{person}{Girish Sastry}, \bibinfo{person}{Amanda Askell}, \bibinfo{person}{Pamela Mishkin}, \bibinfo{person}{Jack Clark}, \bibinfo{person}{Gretchen Krueger}, {and} \bibinfo{person}{Ilya Sutskever}.} \bibinfo{year}{2021}\natexlab{}.
\newblock \showarticletitle{Learning Transferable Visual Models From Natural Language Supervision}.
\newblock \bibinfo{journal}{\emph{arXiv}}.
\newblock


\bibitem[Rajput et~al\mbox{.}(2023)]%
        {rajput2023recommender}
\bibfield{author}{\bibinfo{person}{Shashank Rajput}, \bibinfo{person}{Nikhil Mehta}, \bibinfo{person}{Anima Singh}, \bibinfo{person}{Raghunandan Hulikal~Keshavan}, \bibinfo{person}{Trung Vu}, \bibinfo{person}{Lukasz Heldt}, \bibinfo{person}{Lichan Hong}, \bibinfo{person}{Yi Tay}, \bibinfo{person}{Vinh Tran}, \bibinfo{person}{Jonah Samost}, {et~al\mbox{.}}} \bibinfo{year}{2023}\natexlab{}.
\newblock \showarticletitle{Recommender systems with generative retrieval}.
\newblock \bibinfo{journal}{\emph{Conference on Neural Information Processing Systems (NeurIPS)}} (\bibinfo{year}{2023}).
\newblock


\bibitem[Rendle(2010)]%
        {fm}
\bibfield{author}{\bibinfo{person}{Steffen Rendle}.} \bibinfo{year}{2010}\natexlab{}.
\newblock \showarticletitle{Factorization machines}. In \bibinfo{booktitle}{\emph{IEEE International Conference on Data Mining (ICDM)}}.
\newblock


\bibitem[Sheng et~al\mbox{.}(2024)]%
        {simtier}
\bibfield{author}{\bibinfo{person}{Xiang-Rong Sheng}, \bibinfo{person}{Feifan Yang}, \bibinfo{person}{Litong Gong}, \bibinfo{person}{Biao Wang}, \bibinfo{person}{Zhangming Chan}, \bibinfo{person}{Yujing Zhang}, \bibinfo{person}{Yueyao Cheng}, \bibinfo{person}{Yong-Nan Zhu}, \bibinfo{person}{Tiezheng Ge}, \bibinfo{person}{Han Zhu}, \bibinfo{person}{Yuning Jiang}, \bibinfo{person}{Jian Xu}, {and} \bibinfo{person}{Bo Zheng}.} \bibinfo{year}{2024}\natexlab{}.
\newblock \showarticletitle{Enhancing Taobao Display Advertising with Multimodal Representations: Challenges, Approaches and Insights}. In \bibinfo{booktitle}{\emph{ACM International Conference on Information and Knowledge Management (CIKM)}}.
\newblock


\bibitem[Singh et~al\mbox{.}(2024)]%
        {singh2024better}
\bibfield{author}{\bibinfo{person}{Anima Singh}, \bibinfo{person}{Trung Vu}, \bibinfo{person}{Nikhil Mehta}, \bibinfo{person}{Raghunandan Keshavan}, \bibinfo{person}{Maheswaran Sathiamoorthy}, \bibinfo{person}{Yilin Zheng}, \bibinfo{person}{Lichan Hong}, \bibinfo{person}{Lukasz Heldt}, \bibinfo{person}{Li Wei}, \bibinfo{person}{Devansh Tandon}, {et~al\mbox{.}}} \bibinfo{year}{2024}\natexlab{}.
\newblock \showarticletitle{Better generalization with semantic ids: A case study in ranking for recommendations}. In \bibinfo{booktitle}{\emph{ACM Conference on Recommender Systems (RecSys)}}.
\newblock


\bibitem[Van Den~Oord et~al\mbox{.}(2017)]%
        {van2017neural}
\bibfield{author}{\bibinfo{person}{Aaron Van Den~Oord}, \bibinfo{person}{Oriol Vinyals}, {et~al\mbox{.}}} \bibinfo{year}{2017}\natexlab{}.
\newblock \showarticletitle{Neural discrete representation learning}.
\newblock  (\bibinfo{year}{2017}).
\newblock


\bibitem[Wang et~al\mbox{.}(2021)]%
        {dcn}
\bibfield{author}{\bibinfo{person}{Ruoxi Wang}, \bibinfo{person}{Rakesh Shivanna}, \bibinfo{person}{Derek Cheng}, \bibinfo{person}{Sagar Jain}, \bibinfo{person}{Dong Lin}, \bibinfo{person}{Lichan Hong}, {and} \bibinfo{person}{Ed Chi}.} \bibinfo{year}{2021}\natexlab{}.
\newblock \showarticletitle{Dcn v2: Improved deep \& cross network and practical lessons for web-scale learning to rank systems}. In \bibinfo{booktitle}{\emph{Proceedings of The Web Conference (TheWebConf)}}.
\newblock


\bibitem[Wang et~al\mbox{.}(2022)]%
        {beit-3}
\bibfield{author}{\bibinfo{person}{Wenhui Wang}, \bibinfo{person}{Hangbo Bao}, \bibinfo{person}{Li Dong}, \bibinfo{person}{Johan Bjorck}, \bibinfo{person}{Zhiliang Peng}, \bibinfo{person}{Qiang Liu}, \bibinfo{person}{Kriti Aggarwal}, \bibinfo{person}{Owais~Khan Mohammed}, \bibinfo{person}{Saksham Singhal}, \bibinfo{person}{Subhojit Som}, {and} \bibinfo{person}{Furu Wei}.} \bibinfo{year}{2022}\natexlab{}.
\newblock \showarticletitle{Image as a Foreign Language: BEiT Pretraining for All Vision and Vision-Language Tasks}.
\newblock \bibinfo{journal}{\emph{arXiv}}.
\newblock


\bibitem[Wang et~al\mbox{.}(2024)]%
        {home}
\bibfield{author}{\bibinfo{person}{Xu Wang}, \bibinfo{person}{Jiangxia Cao}, \bibinfo{person}{Zhiyi Fu}, \bibinfo{person}{Kun Gai}, {and} \bibinfo{person}{Guorui Zhou}.} \bibinfo{year}{2024}\natexlab{}.
\newblock \showarticletitle{HoME: Hierarchy of Multi-Gate Experts for Multi-Task Learning at Kuaishou}.
\newblock \bibinfo{journal}{\emph{arXiv}} (\bibinfo{year}{2024}).
\newblock


\bibitem[Wright et~al\mbox{.}(2010)]%
        {wright2010sparse}
\bibfield{author}{\bibinfo{person}{John Wright}, \bibinfo{person}{Yi Ma}, \bibinfo{person}{Julien Mairal}, \bibinfo{person}{Guillermo Sapiro}, \bibinfo{person}{Thomas~S Huang}, {and} \bibinfo{person}{Shuicheng Yan}.} \bibinfo{year}{2010}\natexlab{}.
\newblock \showarticletitle{Sparse representation for computer vision and pattern recognition}.
\newblock \bibinfo{journal}{\emph{Proc. IEEE}} (\bibinfo{year}{2010}).
\newblock


\bibitem[Yang et~al\mbox{.}(2020)]%
        {yang2020large}
\bibfield{author}{\bibinfo{person}{Xiaoyong Yang}, \bibinfo{person}{Yadong Zhu}, \bibinfo{person}{Yi Zhang}, \bibinfo{person}{Xiaobo Wang}, {and} \bibinfo{person}{Quan Yuan}.} \bibinfo{year}{2020}\natexlab{}.
\newblock \showarticletitle{Large scale product graph construction for recommendation in e-commerce}.
\newblock \bibinfo{journal}{\emph{arXiv}} (\bibinfo{year}{2020}).
\newblock


\bibitem[Zhang et~al\mbox{.}(2021)]%
        {lattice}
\bibfield{author}{\bibinfo{person}{Jinghao Zhang}, \bibinfo{person}{Yanqiao Zhu}, \bibinfo{person}{Qiang Liu}, \bibinfo{person}{Shu Wu}, \bibinfo{person}{Shuhui Wang}, {and} \bibinfo{person}{Liang Wang}.} \bibinfo{year}{2021}\natexlab{}.
\newblock \showarticletitle{Mining Latent Structures for Multimedia Recommendation}. In \bibinfo{booktitle}{\emph{ACM International Conference on Multimedia (ACM MM)}}.
\newblock


\bibitem[Zhou et~al\mbox{.}(2019)]%
        {dien}
\bibfield{author}{\bibinfo{person}{Guorui Zhou}, \bibinfo{person}{Na Mou}, \bibinfo{person}{Ying Fan}, \bibinfo{person}{Qi Pi}, \bibinfo{person}{Weijie Bian}, \bibinfo{person}{Chang Zhou}, \bibinfo{person}{Xiaoqiang Zhu}, {and} \bibinfo{person}{Kun Gai}.} \bibinfo{year}{2019}\natexlab{}.
\newblock \showarticletitle{Deep Interest Evolution Network for Click-Through Rate Prediction}. In \bibinfo{booktitle}{\emph{AAAI Conference on Artificial Intelligence (AAAI)}}.
\newblock


\bibitem[Zhou et~al\mbox{.}(2018)]%
        {din}
\bibfield{author}{\bibinfo{person}{Guorui Zhou}, \bibinfo{person}{Xiaoqiang Zhu}, \bibinfo{person}{Chenru Song}, \bibinfo{person}{Ying Fan}, \bibinfo{person}{Han Zhu}, \bibinfo{person}{Xiao Ma}, \bibinfo{person}{Yanghui Yan}, \bibinfo{person}{Junqi Jin}, \bibinfo{person}{Han Li}, {and} \bibinfo{person}{Kun Gai}.} \bibinfo{year}{2018}\natexlab{}.
\newblock \showarticletitle{Deep Interest Network for Click-Through Rate Prediction}. In \bibinfo{booktitle}{\emph{ACM SIGKDD Conference on Knowledge Discovery and Data Mining (KDD)}}.
\newblock


\bibitem[Zhou et~al\mbox{.}(2023)]%
        {bm3}
\bibfield{author}{\bibinfo{person}{Xin Zhou}, \bibinfo{person}{Hongyu Zhou}, \bibinfo{person}{Yong Liu}, \bibinfo{person}{Zhiwei Zeng}, \bibinfo{person}{Chunyan Miao}, \bibinfo{person}{Pengwei Wang}, \bibinfo{person}{Yuan You}, {and} \bibinfo{person}{Feijun Jiang}.} \bibinfo{year}{2023}\natexlab{}.
\newblock \showarticletitle{Bootstrap Latent Representations for Multi-modal Recommendation}. In \bibinfo{booktitle}{\emph{Proceedings of The Web Conference (TheWebConf)}}.
\newblock


\end{thebibliography}
\end{document}